\documentclass[prd,aps,nofootinbib,floatfix,11pt]{revtex4}
\usepackage{amsmath,graphicx,epsfig,amssymb,dsfont,mathtools,}
\usepackage[usenames]{color}
\usepackage{ulem} %% for strike-through
\usepackage{bigstrut}
\usepackage{slashed}
\usepackage{multirow}
\usepackage{subfigure}

\allowdisplaybreaks

\begin{document}\title{Towards the strong decays of light strange four-quark states with $J^P=1^+$}

\author{Ye Xing}
\email{Corresponding author. xingye_guang@cumt.edu.cn}
\affiliation{School of Physics, China University of Mining and Technology, Xuzhou 221000, China}

\author{Zhi-Peng Xing}
\email{Corresponding author. zpxing@nnu.edu.cn}
\affiliation{Department of Physics and Institute of Theoretical Physics, Nanjing Normal University, Nanjing, 210023, China}

\author{Yu-Ji Shi}
\email{Corresponding author. shiyuji@ecust.edu.cn}
\affiliation{School of Physics, East China University of Science and Technology, Shanghai 200237, China}

\begin{abstract}
 This thesis considers the decays of fully light four quark exotic as the main subject. In the context, we study the possible decays from general symmetric analysis: $uu\bar d\bar s$, $dd\bar s\bar u$, $ds\bar u \bar u$, and $su\bar d \bar d$. Using the light quark flavor symmetry, we discuss the decay modes of decuplet and 27 multiplet with $J^P=1^+$. Further more, the phenomenological hadronic molecular approach is employed for the calculation of respective decay modes. Our results show that the total decay widths of $T_{\rho K}$  and $S_{\rho K^*}$ can reach several MeV. The two-body and three-body decay widths are further obtained respectively.
 These are excepted to be fairly valuable supports for future search experiments.
\end{abstract}
\maketitle

%\section{Introduction}

\section{Introduction}
Research on fully light tetraquark has a long tradition, whether in experimental or theoretical research. In 2006, the BaBar collaboration discovered a four-quark candidate Y(2175) with the quantum number $J^P=1^{--}$ in the process $e^+e^-\to \phi f_0(980)$~\cite{BaBar:2006gsq}. This was confirmed by Belle and BES collaborations~\cite{BES:2007sqy,Belle:2008kuo,BESIII:2014ybv}. Recent experiments at collaborations such as Belle, BaBar, and BESIII observed light four-quark candidates, e.g. the X(2239) and X(1835) were seen in the $e^+e^-$ collisions~\cite{BES:2005ega,BESIII:2018ldc}, the X(2100), X(2120) and X(2370) were found in the decay $J/\psi\to \phi\eta\eta'$~\cite{BESIII:2018zbm,BESIII:2010gmv,BESIII:2016qzq}.  Nevertheless, there is still no direct and clear evidence in the current experimental research on four-quark states.  In the theoretical side, the QCD allows existence of four-quark. There have been a number of possible theoretical studies for the fully-light four-quark states, as the compact tetraquark state~\cite{Chen:2008ej,Ke:2018evd,Deng:2010zzd,Ozdem:2022ydv,Huang:2016rro,Chen:2013jra,Chen:2015fwa,Cheng:2024zul,Karliner:2020vsi,Wang:2020xyc}, molecular states~\cite{Dong:2017rmg,Guo:2023igo,Dong:2024fjk,Gutsche:2017twh}, resonance~\cite{MartinezTorres:2008gy,Gomez-Avila:2007pgn} and hybrid states~\cite{Ding:2006ya}. References~\cite{Agaev:2019coa,Azizi:2019ecm} explore light four-quark as a tetraquark candidate using the QCD sum rule. Molecular state candidates were discussed in references~\cite{Guo:2023igo,Dong:2024fjk}.  Within the  hadronic interaction, the ordinary resonant picture is discussed in \cite{MartinezTorres:2008gy}. The possible interpretation based on gluon hybrid state about four-quark candidate has been studied in \cite{Ding:2007pc,Close:2007ny}.

In the present paper, we advocate the molecular state for the $J^P=1^+$ light strange four-quark states under the framework of effective Lagrangian method. The approach has been widely adopted in different hadron system~\cite{Isgur:1989js,Locher:1996nk,Lipkin:1996ny,Li:1996yn,Han:2021azw,Chen:2015iqa}. With advancing research on exotic states, there are growing appeals for the internal structure and dynamics for the those particles. Consequently, we are mainly forced to investigate the two-body and three-body decays, with the addition of possible molecular state as pseudoscalar-vector $PV$ and vector-vector $VV$. We suppose that our analyses of the four-quark state are useful for theoretical study and the future experimental research.

This paper is organized as follows. In Sec.II we briefly discuss the properties of light strange four-quark states. In Sec.III we present an study of the strong decays of four-quark, which is the major topics of the work. We make a conclusion in Sec.IV.

\section{multiplet}
In terms of strictly SU$(3)_f$ flavor symmetry, a fully light four-quark state can be grouped into 8, 10, $\overline {10}$ and 27 states.  Exclusion of state with $q\bar q$ which can electromagnetic decays leaves only decuplet, anti-decuplet, and 27 states. According to the flavor decomposition~\cite{Shi:2020gfp}, the irreducible representation of decuplet $T_{10}$ and anti-decuplet $T_{\overline{10}}$  hold a couple symmetric and antisymmetric indexes form,  %with a pair antisymmetric flavor indexes or $T_{8^{\prime}}$
%\begin{eqnarray*}
$(T_{10})_{ijm}=H_{\{ij}^{kl}\varepsilon_{m\}kl}$ and $(T_{\overline{10}})^{ijm}=H^{\{ij}_{kl}\varepsilon^{m\}kl}$.
%$ (T_{8})^m_n%=\frac{1}{2}\varepsilon_{ijn}\varepsilon^{klm} H^{[ij]}_{[kl]},\quad (T_{8^{\prime}})^m_n
%=H^{\{im\}}_{\{in\}}-\frac{1}{3}\delta^m_n H^{\{ij\}}_{\{ij\}}$.
%\end{eqnarray*}
The %two different reduction modes demonstrate equivalent
flavor irreducible representation with light strange quark, following a exhaustive generalized form as
%\begin{eqnarray}
%T_{8}=\begin{pmatrix}
%T_{(u\bar u-d\bar d)s\bar s}^0+T_{(u\bar u+d\bar d)s\bar s}^0
% &T_{u\bar d s\bar s}^+&T_{u\bar s d\bar d}^+\\
% T_{d\bar u s\bar s}^-&-T_{(u\bar u-d\bar d)s\bar s}^0+T_{(u\bar u+d\bar d)s\bar s}^0&T_{d\bar s u\bar u}^0\\
% T_{s\bar u d\bar d}^-&T_{s\bar d u\bar u}^0&T_{u\bar ud\bar d}^0
% \end{pmatrix},
%\end{eqnarray}
%Namely,
%\begin{eqnarray*}
%&&T_{u\bar s d\bar d}^+=\frac{\sqrt3}{2}S_{\pi \bar K}^{+}-\frac{1}{2}S_{\eta_0 K}^+,\ T_{d\bar s u\bar u}^0=\frac{\sqrt3}{2}S_{\pi \bar K}^{0}-\frac{1}{2}S_{\eta_0 \bar K}^0,\ T_{u\bar d s\bar s}^+=-\frac{1}{\sqrt2}S_{K\bar K}^{+}-\frac{1}{\sqrt{2}}S_{\eta_s \pi}^{+},\\
%&&T_{d\bar u s\bar s}^0=-\frac{1}{\sqrt2}S_{K\bar K}^{0}-\frac{1}{\sqrt2}S_{\eta_s \pi}^{0},\ T_{s\bar d u\bar u}^0=\frac{\sqrt3}{2}S_{K \pi}^{0}-\frac{1}{2}S_{\eta_0 K}^0,\ T_{s\bar u d\bar d}^-=\frac{\sqrt3}{2}S_{K \pi}^{-}-\frac{1}{2}S_{\eta_0 K}^-,\\
%&&T_{(u\bar u-d\bar d)s\bar s}^0=\frac{1}{\sqrt2}S_{K\bar K}^{0}+\frac{1}{\sqrt2}S_{\eta_0\eta_s}^0,\ T_{(u\bar u+d\bar d)s\bar s}^0=\frac{1}{\sqrt2}S_{K\bar K}^{0}+\frac{1}{\sqrt2}S_{\eta_0\eta_s}^0,\ T_{u\bar ud\bar d}^0=\frac{\sqrt3}{2}S_{\pi\pi}^0+\frac{1}{2}S_{\eta_0\eta_0}^0.
%\end{eqnarray*}
\begin{eqnarray*}
&&(T_{10})_{\{111\}}=\frac{1}{2}(uu\bar d\bar s-uu\bar s\bar d)=T_{\rho  \bar K}^{++},\  (T_{10})_{\{222\}}=\frac{1}{2}(dd\bar s\bar u-dd\bar u\bar s)=T_{\rho \bar K}^{-},\\
&&(T_{\overline{10}})^{\{111\}}=\frac{1}{2}(ds\bar u\bar u-sd\bar u\bar u)=T_{\rho  K}^{--},\  (T_{\overline{10}})^{\{222\}}=\frac{1}{2}(us\bar d\bar d-su\bar d\bar d)=T_{\rho K}^{+}.
\end{eqnarray*}
For the light strange four-quark 27 state aspect, the irreducible representations $T_{27}$  posses two pairs symmetric indexes,
%\begin{eqnarray*}
$(T_{27})_{\{kl\}}^{\{ij\}}=(H_{\{kl\}}^{\{ij\}})_{traceless}$.
%\end{eqnarray*}
Naturally, $T_{27}$ is obtained as follows,
%\begin{eqnarray*}
%&&(T_{10})_{\{111\}}=T_{\pi \bar K}^{++},\ (T_{10})_{\{112\}}=-\frac{1}{2}T_{\pi \bar K}^{+}+\frac{\sqrt{3}}{2}T_{\eta_0 \bar K}^+,\ (T_{10})_{\{122\}}=-\frac{1}{2}T_{\pi \bar K}^{0}+\frac{\sqrt{3}}{2}T_{\eta_0 \bar K}^0,\\
%&&(T_{10})_{\{113\}}=\frac{1}{2}(T_{\eta_0 \pi}^{+}-T_{\pi \pi}^{+})+\frac{1}{2}(T_{\eta_s \pi}^+-T_{K\bar K}^+),\ (T_{10})_{\{123\}}=\frac{1}{2}(T_{\eta_0 \pi}^{0}-T_{\pi \pi}^{0})+\frac{1}{2}(T_{\eta_s \eta_0}^0-T_{K\bar K}^0),\\
%&&(T_{10})_{\{223\}}=\frac{1}{2}(T_{\eta_0 \pi}^{-}-T_{\pi \pi}^{-})+\frac{1}{2}(T_{\eta_s \pi}^--T_{K\bar K}^-),\ (T_{10})_{\{133\}}=\frac{\sqrt2}{4}T_{\eta_0  K}^{0}-\frac{\sqrt{6}}{4}T_{K \pi}^0+\frac{\sqrt2}{2}T_{\eta_s K}^0,\\
%&&(T_{10})_{\{233\}}=\frac{\sqrt2}{4}T_{\eta_0  K}^{-}-\frac{\sqrt{6}}{4}T_{K \pi}^-+\frac{\sqrt2}{2}T_{\eta_s K}^-,\  (T_{10})_{222}=T_{\pi \bar K}^{-},\ (T_{10})_{333}=T_{K  K}^{-}.
%\end{eqnarray*}
\begin{eqnarray*}
&&(T_{27})_{11}^{\{23\}}=\frac{1}{2}(ds\bar u\bar u+sd\bar u\bar u)=S_{\rho K^*}^{--},\ (T_{27})_{\{13\}}^{22}=\frac{1}{2}(dd\bar u\bar s+dd\bar s\bar u)=S_{\rho \bar K^*}^{-},\\
&&(T_{27})_{22}^{\{13\}}=\frac{1}{2}(us\bar d\bar d+su\bar d\bar d)=S_{\rho K^*}^{+},\ (T_{27})_{\{23\}}^{11}=\frac{1}{2}(uu\bar d\bar s+uu\bar s\bar d)=S_{\rho \bar K^*}^{++},
\end{eqnarray*}
To our knowledge, the S-wave hadronic multiplet are illuminated in the \textbf{flavor} $\otimes$ \textbf{color} $\otimes$ \textbf{spin} space. Depending on the prior flavor structure, it then turns out two possible constituents respectively for decuplet($T_{10}$ or $T_{\overline{10}}$) and twenty-seven state($T_{27}$) in respect of diquark scheme,
\begin{eqnarray}\label{eq:decomp}
%&T_8:\quad& | [qq]_{\bar 3_\textbf{f}}[\bar q\bar q]_{3_\textbf{f}},8_\textbf{f}\rangle\otimes| [qq]_{\bar 3_\textbf{c}}[\bar q\bar q]_{3_\textbf{c}},1_\textbf{c}\rangle\otimes| [qq]_{0_\textbf{s}},[\bar q\bar q]_{0_\textbf{s}},0_\textbf{s}\rangle,\\
%    && | [qq]_{\bar 3_\textbf{f}}[\bar q\bar q]_{3_\textbf{f}},8_\textbf{f}\rangle\otimes| \{qq\}_{6_\textbf{c}}\{\bar q\bar q\}_{\bar 6_\textbf{c}},1_\textbf{c}\rangle\otimes| \{qq\}_{1_\textbf{s}},\{\bar q\bar q\}_{1_\textbf{s}},(0/1/2)_\textbf{s}\rangle,\nonumber\\
&T_{27}:\quad& | \{qq\}_{6_\textbf{f}}\{\bar q\bar q\}_{\bar 6_\textbf{f}},27_\textbf{f}\rangle\otimes| [qq]_{\bar 3_\textbf{c}}[\bar q\bar q]_{3_\textbf{c}},1_\textbf{c}\rangle\otimes| \{qq\}_{1_\textbf{s}},\{\bar q\bar q\}_{1_\textbf{s}},(0/1/2)_\textbf{s}\rangle,\nonumber\\
   % && | \{qq\}_{6_\textbf{f}}\{\bar q\bar q\}_{\bar 6_\textbf{f}},27_\textbf{f}\rangle\otimes| \{qq\}_{6_\textbf{c}}\{\bar q\bar q\}_{\bar 6_\textbf{c}},1_\textbf{c}\rangle\otimes| [qq]_{0_\textbf{s}},[\bar q\bar q]_{0_\textbf{s}},0_\textbf{s}\rangle,\nonumber\\
&T_{10}:\quad& | [qq]_{\bar 3_\textbf{f}}\{\bar q\bar q\}_{\bar 6_\textbf{f}},10_\textbf{f}\rangle\otimes| [qq]_{\bar 3_\textbf{c}}[\bar q\bar q]_{3_\textbf{c}},1_\textbf{c}\rangle\otimes| [qq]_{0_\textbf{s}},\{\bar q\bar q\}_{1_\textbf{s}},1_\textbf{s}\rangle.
    %&& | [qq]_{\bar 3_\textbf{f}}\{\bar q\bar q\}_{\bar 6_\textbf{f}},10_\textbf{f}\rangle\otimes| \{qq\}_{6_\textbf{c}}\{\bar q\bar q\}_{\bar 6_\textbf{c}},1_\textbf{c}\rangle\otimes| \{qq\}_{1_\textbf{s}},[\bar q\bar q]_{0_\textbf{s}},1_\textbf{s}\rangle.
\end{eqnarray}
Here, we identify the four-quark $|(qq)_{\textbf{j}_1}(\bar q\bar q)_{\textbf{j}_2},J\rangle$ individually in flavor(\textbf{f}), colour(\textbf{c}) and spin(\textbf{s}) space. $j_1(j_2)$ is the quantum number of diquark(antidiquark) in different subspace, and $J$ is the total quantum number of four-quark in the subspace.
%The diquark with color-$\bar 3$ and spin-$0$ is known as ``good'' diquark, since it matches with a more attractive potential. Another one with color-$\bar 3$ and spin-$1$ is proposed to be ``bad'' diquark.
$T_{27}$ has a pair of symmetric diquarks in flavor subspace %$| [qq]_{\bar 3_\textbf{f}}[\bar q\bar q]_{3_\textbf{f}},8_\textbf{f}\rangle$, while $T_{8^{\prime}}$ owns two symmetric diquarks
$| \{qq\}_{6_\textbf{f}}\{\bar q\bar q\}_{\bar 6_\textbf{f}},27_\textbf{f}\rangle$ and color-($3,\bar 3$) diquarks in color subspace $| [qq]_{\bar 3_\textbf{c}}[\bar q\bar q]_{3_\textbf{c}},1_\textbf{c}\rangle$, then the spin of %color-($3,\bar 3$)
$T_{27}$ can be spin-0,1,2.
%Under exact SU(3) light flavor symmetry, the two octet $T_8, T_{8^{\prime}}$ with different flavor substructures can not be mixed with each other.
% thus we wound discuss them separately in the following.
The flavor decuplet $T_{10}$ includes a couple of antisymmetric and symmetric diquarks, only leading to spin-$1$ S-wave state. The diquark with color-$\bar 6$ and spin-$0$ is a roughly potential possibility, however recent Lattice QCD\cite{Yeo:2024chk} show unfavourable result for the diquark.
%Remembering that all assignments give priority to the flavor eigenstate.
Then in the work, referring to the possible constituents, it is available to arrange the S-wave strange particle states with $J^P=0^+, 1^+, 2^+$ to 27 states and $J^P=1^+$ to decuplet.

In the framework of color-spin interaction model, we roughly discuss the mass spectrum, by means of the interaction Hamiltonian given as~\cite{DeRujula:1975qlm}
\begin{eqnarray}
\mathcal{H}_i=\sum_{i<j}\Big(-\frac{3}{8}\Big) \frac{C^{ij}}{m_i m_j} \lambda_i\cdot\lambda_j s_i\cdot s_j,
\end{eqnarray}
here, $\lambda_i$ is the Gell-Mann matrix and $s_i=\sigma_i/2$ is the quark spin operator. A overall strength $C^{ij}$ can be extracted from conventional meson and baryon system\cite{Xing:2018bqt}. We found that the mass spectrum of $1^+$ state,
\begin{eqnarray}
m(T_{\rho K})=1.375\pm0.04\ \text{GeV},\quad m(S_{\rho K^*})=1.439\pm0.05\ \text{GeV},\quad J^P=1^+.
\end{eqnarray}
The decuplet four-quark state $1^+$ lie in the mass  1.375 GeV near the threshold of $K\rho$, the 27 state lie in the 1.439 GeV near the thresholds of   $K^*\rho$, then we name the $T_{10}$ as $T_{\rho K}$ and $T_{27}$ as $S_{\rho K^*}$.  Under the consideration, the four-quark state $T_{\rho K}$  can decay strongly into $\pi K^*$, and $S_{\rho K^*}$ may decay into $K^* \pi,\rho K$.

\section{Decay Modes}
We now turn to the hadronic decays of the light strange four-quark states. The concerned four-quark states are advocated as $ns \bar n\bar n$ with $J^P=1^+$. A processable framework is the SU(3) symmetry analysis\cite{He:2021qnc,Xing:2024nvg,Li:2023kcl}, as well as effective Lagrangian approach\cite{Li:2024fmg}.

In our considering, the light four-quark can undergo decay against strong interaction. It is directly to write the Hamiltonian, the $T_{27}$s and $T_{10}$s can decay into a pseudoscalar($M$) and vector meson($V$), while the Hamiltonian of $T_{10}$ decays into two pseudoscalar($M$) or vector mesons($V$) are forbidden for the SU(3) symmetry,
\begin{eqnarray}
\mathcal{H}_{10}&=&a_1(T_{10})^{\alpha ij} \varepsilon_{\alpha kl} M^k_i V^l_j+b_1(T_{10})^{\alpha ij} \varepsilon_{\alpha kl} M^k_m M^l_i M^m_j,\\
\mathcal{H}_{27}&=&
%a_1(T_{27})^{\{ij\}}_{\{kl\}} M^k_i M^l_j+
a_1'(T_{27})^{\{ij\}}_{\{kl\}} V^k_i M^l_j+b_1' (T_{27})^{\{ij\}}_{\{kl\}} M^k_i M^l_m M^m_j,
%+(T_{10})^{\alpha ij} \varepsilon_{\alpha kl} V^k_m M^l_i M^m_j\\
%+(T_{10})^{\alpha ij} \varepsilon_{\alpha kl} V^k_i M^l_m M^m_j+(T_{10})^{\alpha ij} \varepsilon_{\alpha lm} V^k_i M^l_j M^m_k,
\end{eqnarray}
accordingly, the decay processes of 10 state($T_{\rho K}$) and 27 state($S_{\rho K^*}$) are exhibited as,
\begin{eqnarray*}
% \nonumber to remove numbering (before each equation)T_{(u\bar u-d\bar d) s\bar s}^0)=m(T_{u\bar ds\bar s}^+)=m(T_{d\bar us\bar s}^-)=m(T_{s\bar s u\bar u}^0
T_{\rho K} \rightarrow K^*\pi,\  K\pi\pi,\quad
%  T_{u\bar s d\bar d}^+\to \bar K^0 \pi^-, K^- \pi^0, \eta K^-,\\
%  T_{s\bar u d\bar d}^-\to K^0 \pi^+, K^+ \pi^0,  \eta K^+,\\
%  T_{s\bar d u\bar u}^0\to K^+ \pi^-, K^0 \pi^0, \eta K^0,\\
%  T_{d\bar s u\bar u}^0\to K^- \pi^+, \bar K^0 \pi^0, , \eta \bar K^0,
%  T_{(u\bar u-d\bar d) s\bar s}^0/T_{u\bar ds\bar s}^+/T_{d\bar us\bar s}^-/T_{s\bar s u\bar u}^0\rightarrow \pi\eta, K\bar K,
%T_{\pi\pi}\rightarrow \pi\rho, \quad
S_{\rho K^*}\rightarrow  K\rho,K^*\pi,\  K\pi \pi, KK K.
%T_{K K}\rightarrow KK^*.
\end{eqnarray*}
Expanding the Hamiltonian, the relation between different channels can be deduced, which are collected into tab~\ref{tab:1} and tab~\ref{tab:2},
\begin{table}
	\caption{The decay relation of different decuplet $T_{10}$ channels.}\label{tab:2}\begin{tabular}{clclc|c|c|c}\hline\hline
		$\Gamma(T^{--}_{\rho K}\to \pi^{-}K^{*-} )$&$=\Gamma(T^{+}_{\rho K}\to \pi^{+}K^{*0})$&$\Gamma(T^{--}_{\rho K}\to \pi^{-}\pi^- K^0 )$&$=4\Gamma(T^{--}_{\rho K}\to \pi^{-}\pi^0 K^- )$\\
		&$=\Gamma(T^{++}_{\rho \bar K}\to \pi^{+}K^{*+} )$& &$=\Gamma(T^{-}_{\rho \bar K}\to \pi^{-}\pi^- K^+ )$\\
		&$=\Gamma(T^{-}_{\rho \bar K}\to \pi^{-}K^{*0})$&&$=4\Gamma(T^{-}_{\rho \bar K}\to \pi^{0}\pi^- K^0 )$\\
		  &&&$=4\Gamma(T^{++}_{\rho \bar K}\to \pi^{+}\pi^0 K^+ )$\\
		&&&$=\Gamma(T^{++}_{\rho \bar K}\to \pi^{+}\pi^+ K^0 )$\\
        &&&$=\Gamma(T^{+}_{\rho K}\to \pi^{+}\pi^+K^- )$\\
        &&&$=4\Gamma(T^{+}_{\rho K}\to \pi^0\pi^+K^0 )$\\

\hline
\end{tabular}
\end{table}
\begin{table}
	\caption{The decay relation of different twenty-seven $T_{27}$ channels.}\label{tab:1}\begin{tabular}{clclc|c|c|c}\hline\hline
		$\Gamma(S^{--}_{\rho K^*}\to\pi^- K^{*-})$&$=\Gamma(S^{--}_{\rho K^*}\to K^- \rho^- )$&$\Gamma(S^{--}_{\rho K^*}\to\pi^{0}\pi^- K^-)$&$=\frac{1}{4}\Gamma(S^{--}_{\rho K^*}\to \pi^{-}\pi^- \overline{K}^0 )$ \\
		&$=\Gamma(S^{++}_{\rho \bar{K}^*}\to K^+ \rho^{+})$&&$=\frac{1}{4}\Gamma(S^{--}_{\rho K^*}\to K^0 K^{-} K^{-} )$ \\
		&$=\Gamma(S^{++}_{\rho \bar{K}^*}\to \pi^{+}K^{*+} )$&&$=\frac{1}{4}\Gamma(S^{+}_{\rho K^*}\to \pi^{+}\pi^+ K^- )$\\
		&$=\Gamma(S^{-}_{\rho \bar{K}^*}\to K^0 \rho^{-})$&&$=\Gamma(S^{+}_{\rho K^*}\to \pi^{+}\pi^0 \overline{K}^0 )$\\
		&$=\Gamma(S^{+}_{\rho K^*}\to \pi^+ \overline{K}^{*0} )$&&$=\frac{1}{4}\Gamma(S^{+}_{\rho K^*}\to K^{+}\overline{K}^0 \overline{K}^0 )$\\
		&$=\Gamma(S^{+}_{\rho K^*}\to \bar{K}^0 \rho^{+})$&&$=\Gamma(S^{-}_{\rho \bar{K}^*}\to \pi^{0}\pi^- K^0 )$\\
		&$=\Gamma(S^{-}_{\rho \bar{K}^*}\to \pi^{-} \overline{K}^{*0} )$&&$=\frac{1}{4}\Gamma(S^{-}_{\rho \bar{K}^*}\to \pi^{-}\pi^- K^+ )$\\
		&&&$=\frac{1}{4}\Gamma(S^{-}_{\rho \bar{K}^*}\to K^{0}K^{0} K^- )$\\
		&&&$=\frac{1}{4}\Gamma(S^{++}_{\rho \bar{K}^*}\to \pi^{+}\pi^+ K^0 )$\\
		&&&$=\Gamma(S^{++}_{\rho \bar{K}^*}\to \pi^{+}\pi^0 K^+ )$\\
		&&&$=\frac{1}{4}\Gamma(S^{++}_{\rho \bar{K}^*}\to K^{+}K^+ \overline{K}^0 )$\\
\hline
	\end{tabular}
\end{table}
\subsection{four-quark molecular state($1^+$) coupled by $PV$ or $VV$}
The decuplet four-quark state $1^+$ lie in the mass of $K\rho$, the 27 state lie in the mass of $K^*\rho$, then it is possible to be a pseudoscalar-vector $PV$ molecular state for decuplet which couple by $K\rho$, and a vector-vector $VV$ molecular state for 27 state which couple by $K^*\rho$. The couplings can be achieved by the mass operator\cite{Gutsche:2010jf,Dong:2008gb}.
%\begin{eqnarray}
%L_{int}=g_{TPP} T(x) \int d y \Phi_T(y^2) P_1^T(x+\omega_1 y) P_2(x-\omega_2 y)+H.c.
%\end{eqnarray}
\begin{eqnarray}
&&\mathcal{L}_{int}^{TPV}= g_{TPV} \bar{T}_{\mu}(x) \int d y \Phi_T(y^2) P(x+\omega_1 y) V^{\mu}(x-\omega_2 y),\\
&&\mathcal{L}_{int}^{TVV}=i g_{TVV} \varepsilon_{\mu\nu\alpha\beta}\partial^{\mu}\bar{T}^{\nu}(x) \int d y \Phi_T(y^2) V_1^{\alpha}(x+\omega_1 y) V_2^{\beta}(x-\omega_2 y),
\end{eqnarray}
where $\omega_1=m_1/(m_1+m_2)$, $\omega_2=m_2/(m_1+m_2)$.
To describe the distribution of the constituents, we similarly introduce the correlation function $\phi(y^2)$, in order to remove the ultraviolet divergence in Feynman diagrams. The Fourier transition in momentum space $\widetilde{\Phi}(-p^2)$ as
\begin{eqnarray}
\Phi(y^2)=\int \frac{d^4 p}{(2\pi)^4} e^{-ipy} \widetilde{\Phi}(-p^2),
\end{eqnarray}
here, $\widetilde{\Phi}(-p^2)$ is chosen as a Gaussian-like form $\widetilde{\Phi}(-p^2)=exp(p^2/\Lambda_E^2)$, where $\Lambda_E$ is the model parameter, which has dimension of mass and defines the scale for the distribution of the constituents. The coupling constant is determined using the Weinberg-Salam compositeness condition, which means that the probability to find the dressed bound state as a bare state is equal to zero $Z=1-\mathcal{M}^{\prime}(m_T^2)=0$,
%\begin{eqnarray}
%Z=1-\mathcal{M}^{\prime}(m_T^2)=0.
%\end{eqnarray}
Thus the amplitudes, reading:
\begin{eqnarray}
&&\mathcal{M}^{\mu\nu}_a=-g_{{\scriptscriptstyle TPV}}^2 \int\!\! \frac{d^4 q}{(2\pi)^4 i}\widetilde{\Phi}^2(-(q-p \omega_2)^2) \frac{-g^{\mu\nu}+q^{\mu}q^{\nu}/m_V^2}{(q^2-m_V^2)((p-q)^2-m_P^2)},\\
&&\mathcal{M}^{\mu\nu}_b=g_{{\scriptscriptstyle TVV}}^2 \int\!\! \frac{d^4 q}{(2\pi)^4 i}\widetilde{\Phi}^2(-(q-p \omega_2)^2) \frac{\varepsilon^{\mu\alpha\beta\rho}\varepsilon^{\nu\alpha'\beta'\rho'}p_{\alpha}p_{\alpha'}(g_{\beta\beta'}-\frac{q_{\beta}q_{\beta'}}{m_{V_1}^2})(g_{\rho\rho'}-\frac{(p-q)_{\rho}(p-q)_{\rho'}}{m_{V_2}^2})}{(q^2-m_{V_1}^2)((p-q)^2-m_{V_2}^2)}.
\end{eqnarray}
%then
%\begin{eqnarray}
%&\frac{1}{g^2}=&-\frac{1}{8\pi^2\Lambda^2}\int^1_0 d \alpha \int_0^{\infty} d\beta \frac{\beta^2}{(1+\beta)^3}(-\omega_2^2-\alpha+2\omega_2 \alpha-(1-\alpha)\alpha\beta) e^{-2z^{\prime}/\Lambda^2},\\
%&&z^{\prime}=-M^2(\alpha\beta+\omega_2^2)+\frac{M^2(\omega_2+\alpha\beta)^2}{1+\beta}+m_2^2 \alpha\beta+m_1^2 (1-\alpha)\beta,\nonumber
%\end{eqnarray}
For the spin-1 state, the self energy can be divided into transverse and longitudinal parts,
\begin{eqnarray}
\mathcal{M}^{\mu\nu}=g^{\mu\nu}_{\bot} \mathcal{M}^T(p^2)+\frac{p^{\mu}p^{\nu}}{p^2} \mathcal{M}^{L}(p^2),
\end{eqnarray}
with $g_{\bot}^{\mu\nu}=g^{\mu\nu}-\frac{p^{\mu}p^{\nu}}{p^2}$, The compositeness condition can then be estimated from the transverse part of the self energy,
\begin{eqnarray}
Z=1-\frac{d\mathcal{M}^T(p^2)}{dp^2}\Big|_{p^2=m_T^2}=0,
\end{eqnarray}
then
\begin{eqnarray}
&\frac{1}{g_{TPV}^2}=&\frac{1}{16\pi^2}\int^1_0\!\! d \alpha \int_0^{\infty}\!\! d\beta \frac{\beta}{(1+\beta)^4}(\frac{2 \beta(1+\beta)}{\Lambda_E^2}+\frac{\beta}{2m_1^2})(\omega_2^2+\alpha-2\omega_2 \alpha+(1-\alpha)\alpha\beta) e^{\frac{-2z^{\prime}}{\Lambda_E^2}},\\
&\frac{1}{g_{TVV}^2}=&\frac{\partial}{\partial p^2}\frac{1}{8\pi^2}\int^1_0\!\! d \alpha \int_0^{\infty}\!\! d\beta \frac{\beta}{(1+\beta)^3}\Big(\frac{m_1^2+m_2^2}{3m_1^2m_2^2}\big(\frac{(\alpha\beta+\omega_2)^2m_T^4}{1+\beta}-(\alpha\beta+\omega_2)^2-\Lambda_E^2 m_T^2\nonumber\\&&+\frac{\Lambda_E^2}{4}(1+\beta)\big)-m_T^2(1+\beta)\Big) e^{-2z^{\prime}/\Lambda_E^2},\\
&&z^{\prime}=-m_T^2(\alpha\beta+\omega_2^2)+\frac{m_T^2(\omega_2+\alpha\beta)^2}{1+\beta}+m_2^2 \alpha\beta+m_1^2 (1-\alpha)\beta.
\end{eqnarray}
In order to evaluate the decay modes, we need the structure of the involved interaction vertices which can be described by means of the following effective Lagrangian\cite{Shen:2019evi},
\begin{eqnarray}
&\mathcal{L}_{\scriptscriptstyle{\tiny PPV}}&=g_{{\scriptscriptstyle PPV}}\ \big(\phi_P(x) \partial_{\mu} \phi_P(x)- \partial_{\mu} \phi_P(x) \phi_P(x)\big) \phi_V^{\mu}(x),\\
&\mathcal{L}_{{\scriptscriptstyle VVP}}&=g_{{\scriptscriptstyle VVP}}\ i\varepsilon_{\mu\nu\alpha\beta}\partial^{\mu} \phi^{\nu}_V(x) \partial^{\alpha} \phi^{\beta}_V(x)\phi_P(x),\\
&\mathcal{L}_{{\scriptscriptstyle VVV}}&=g_{{\scriptscriptstyle VVV}}\ i\big\langle(\partial_{\mu}\phi_{V\nu}(x)-\partial_{\nu}\phi_{V\mu}(x))\phi_{V}^{\mu}(x)\phi^{\nu}_V(x)\big\rangle.
\end{eqnarray}
The coupling $g_{\pi\pi\rho}=6.04, g_{\pi K K^*}=-3.02, g_{\pi \rho K^*}=-7.93, g_{\rho\rho\pi}=-6.45$~\cite{Lin:2018kcc},
%The  amplitude of three body
%\begin{eqnarray}
%&&M_d=g_{{\scriptscriptstyle \pi \pi\rho}} g_{{\scriptscriptstyle  TPV} }(p_2^{\mu}+p_1^{\mu})\epsilon^{\alpha}(p)\frac{i(-g_{\mu\alpha}+\frac{q_{\mu}q_{\alpha}}{m_{\rho}^2})}{q^2-m_{\rho}^2} .
%\end{eqnarray}
Here, we only show the leading order amplitudes of decuplet two-body and three-body decays,
\begin{eqnarray}
%&&M_a=\int \frac{d^4q}{(2\pi)^4} g_{\pi KK^*}g_{\pi\pi\rho}g_{TPV}(p_2^{\mu}-q^{\mu})(q^{\nu}-q_1^{\nu})\epsilon_{\nu}^*(p_1)\epsilon^{\alpha}(p)\frac{i(-g_{\mu\alpha}+\frac{q_{2\mu}q_{2\alpha}}{m_{\rho}^2})}{(q_2^2-m_{\rho}^2)(q^2-m_q^2)(q_1^2-m_K^2)} ,\nonumber\\
&&\mathcal{M}_a=\int \frac{d^4q}{(2\pi)^4}  g_{{\scriptscriptstyle \pi\! K\!K^{\!*}}} g_{{\scriptscriptstyle \pi\pi\rho}}g_{{\scriptscriptstyle TPV}}(-p_{2}-q)_{\mu}(q+q_1)^{\nu}\epsilon_{\nu}^*(p_1)\epsilon^{\alpha}(p){S}^{\mu}_{\boldsymbol{\rho}\alpha}(q_2)S_{\boldsymbol{\pi}}(q)S_{\boldsymbol{K}}(q_1) ,\nonumber\\
%&&M_b=\int \frac{d^4q}{(2\pi)^4} g_{K\rho K^*}g_{\pi\rho\rho} g_{TPV} \varepsilon_{\mu\nu\alpha\beta}\varepsilon_{\mu'\nu'\alpha'\beta'}p_1^{\mu'}q_2^{\mu}q^{\alpha}q^{\alpha'}\epsilon_{m}(p)\epsilon^{*\nu'}(p_1)
%\frac{i(-g^{\beta\beta'}+\frac{q^{\beta}q^{\beta'}}{m_q^2})(-g^{\nu m}+\frac{q_2^{\nu}q_2^{m}}{m_{\rho}^2})}{(q_2^2-m_{\rho}^2)(q^2-m_q^2)(q_1^2-m_K^2)},\nonumber\\
&&\mathcal{M}_b=\int \frac{d^4q}{(2\pi)^4} g_{{\scriptscriptstyle \rho\! K^{\!*}\!K}}g_{{\scriptscriptstyle \rho\rho\pi}} g_{{\scriptscriptstyle TPV}} \varepsilon_{\mu\nu\alpha\beta}\varepsilon_{\nu'\mu'\alpha'\beta'}p_1^{\mu'}q_2^{\mu}q^{\alpha}q^{\alpha'}\epsilon_{m}(p)\epsilon^{*\nu'}(p_1)
S_{\boldsymbol{\rho}}^{m\nu}(q_2)S_{\boldsymbol{\rho}}^{\beta\beta'}(q)S_{\boldsymbol{K}}(q_1),\nonumber\\
%&&M_c=\int \frac{d^4q}{(2\pi)^4} g_{\pi K K^*}g_{\rho K^* K^*} g_{TPV} p_1^{\alpha}( 2p_2^{\mu}g^{\beta}_{\nu}+q_{\nu}g^{\mu\beta}-{p_2}_{\nu}g^{\mu\beta})\epsilon_{m}(p)\epsilon^*_{\beta}(p_2)
%\frac{-(-g_{\mu\alpha}+\frac{q_{\mu}q_{\alpha}}{m_q^2})(-g^{\nu m}+\frac{{q_2}^{\nu}q_2^{m}}{m_{\rho}^2})}{(q_2^2-m_{\rho}^2)(q^2-m_q^2)(q_1^2-m_K^2)}.
&&\mathcal{M}_c=\int \frac{id^4q}{(2\pi)^4} g_{{\scriptscriptstyle \pi\! K\! K^{\!*}}}g_{{\scriptscriptstyle K^{\!*}\! K^{\!*}\!\rho}} g_{{\scriptscriptstyle TPV}}( p_{1}+q_{1})_{\mu}( p_{2}+q)_{\nu}\epsilon^{*}_{\alpha}(p_2)\epsilon_{\beta}(p)
S_{\boldsymbol{K^*}}^{\mu\alpha}(q)S_{\boldsymbol{\rho}}^{\nu\beta}(q_2)S_{\mathbf{K}}(q_1),\nonumber\\
%,\nonumber\\
%%&&M_d=g_{\pi \pi\rho}g_{TPV}(p_2^{\mu}+p_1^{\mu})\epsilon^{\alpha}(p)\frac{i(-g_{\mu\alpha}+\frac{q_{\mu}q_{\alpha}}{q^2})}{(q^2-m_q^2)} .
&&M_d=g_{{\scriptscriptstyle \pi \pi\rho}} g_{{\scriptscriptstyle  TPV} }(p_2+p_1)^{\mu}\epsilon^{\alpha}(p)\frac{i(-g_{\mu\alpha}+\frac{q_{\mu}q_{\alpha}}{m_{\rho}^2})}{q^2-m_{\rho}^2},
\end{eqnarray}
with propagator of vector and pseudoscalar meson given as
\begin{eqnarray}
S_{\mathbb{V}}^{\mu\nu}(q_1)=\frac{i(-g^{\mu \nu}+\frac{q_1^{\mu}q_1^{\nu}}{m_{\mathbb{V}}^2})}{q_1^2-m_{V}^2},\ S_{\mathbb{P}}(q_2)=\frac{i}{q_2^2-m_{\mathbb{P}}^2}.
\end{eqnarray}
%The decay width
%\begin{eqnarray*}
%\Gamma(T\to \mathcal{M}_1+\mathcal{M}_2)=\frac{|\mathbf{P_1}|}{8\pi m_{T}^2}|M|^2,
%\end{eqnarray*}
%where $\mathbf{P_1}$ is the magnitude of three momentum of the final state meson, given as $|\mathbf{P_1}|=\frac{1}{2m_{B_c}}\sqrt{\lambda(m_{B_c}^2,m^2_{\mathcal{B}_{cc}},m^2_{\overline {\mathcal{B}}_{\bar c}})}$; $\lambda(a,b,c)=a^2+b^2+c^2-2ab-2bc-2ac$.
The amplitudes of 27 state two-body decays are deduced as,
\begin{eqnarray}
%&&\mathcal{M}_a^{PP}=\int \frac{d^4q}{(2\pi)^4}  g_{{\scriptscriptstyle \pi\! K\!K^{\!*}}} g_{{\scriptscriptstyle \pi\pi\rho}}g_{{\scriptscriptstyle TVV}}(p_{2}+q)_{\mu'}(q-p_1)_{\nu'}p_{\mu}\varepsilon^{\mu\nu\alpha\beta} \epsilon_{\nu}(p){S}^{\mu'}_{\boldsymbol{\rho}\beta}(q_2)S^{\nu'}_{\boldsymbol{K^*}\alpha}(q_1)S_{\boldsymbol{\pi}}(q) ,\nonumber\\
%&&M_a=\int \frac{d^4q}{i(2\pi)^4} g_{K^* K^* \pi}g_{\pi\pi\rho}g_{TVV}(q^{\mu}-p_2^{\mu})(q^{\nu}-q_1^{\nu})p^{\alpha}p_1^{\mu'}q_1^{\rho}\varepsilon_{\alpha\beta\gamma\delta}\varepsilon_{\mu'\nu\rho\sigma}\epsilon^{\beta}(p)\epsilon^{*\sigma}(p_1)
%\frac{(-g_{\mu}^{\gamma}+\frac{q_{2\mu}q_2^{\gamma}}{m_{\rho}^2})
%(-g^{\delta\nu}+\frac{q_{2}^{\delta}q_2^{\nu}}{m_{K^*}^2})}{(q_2^2-m_{\rho}^2)(q^2-m_q^2)(q_1^2-m_{K^*}^2)} ,\nonumber\\
&&\mathcal{M}_a=\int \frac{id^4q}{(2\pi)^4} g_{{\scriptscriptstyle K^{\!*}\! K^{\!*}\! \pi}}g_{{\scriptscriptstyle \pi\pi\rho}}g_{{\scriptscriptstyle TVV}}(p_2+q)^{\mu}p^{\alpha}p_1^{\mu'}q_1^{\rho}\varepsilon_{\alpha\beta\gamma\delta}\varepsilon_{\mu'\nu\rho\sigma}\epsilon^{\beta}(p)\epsilon^{*\sigma}(p_1)
{S_{\boldsymbol{\rho}}}^{\gamma}_{\mu}(q_2)S_{\boldsymbol{\pi}}(q)S^{\nu\delta}_{\boldsymbol{K^*}}(q_1) ,\nonumber\\
%&&M_b=\int \frac{d^4q}{i(2\pi)^4} g_{K^* K^*\rho}g_{\pi\rho\rho} g_{TVV} \varepsilon_{\alpha\beta\gamma\delta}\varepsilon^{\mu'm\rho\sigma}p^{\alpha}q_{2\mu'}q_{\rho}(q_{\mu}g_{\nu}^{\nu'}-q_{\nu}g_{\mu}^{\nu'})\epsilon^{*\nu}(p_1)\epsilon^{\beta}(p)
%\frac{(-g_{\nu'\sigma}+\frac{q_{\nu'}q^{\sigma}}{m_q^2})(-g^{\mu \delta}+\frac{q_1^{\mu}q_1^{\delta}}{m_{K^*}^2})(-g^{\gamma}_{ m}+\frac{q_2^{\gamma}q_{2m}}{m_{\rho}^2})}{(q_2^2-m_{\rho}^2)(q^2-m_q^2)(q_1^2-m_{K^*}^2)},\nonumber\\
%&&\mathcal{M}_b^{PP}=\int \frac{d^4q}{(2\pi)^4} g_{{\scriptscriptstyle K^{\!*}\! \rho\pi}}g_{{\scriptscriptstyle \rho \rho\pi}} g_{{\scriptscriptstyle TVV}} p^{\mu}p_2^{\mu'}{q}^{\alpha'}q^{\bar \mu}p_1^{\bar \alpha}\varepsilon_{\mu\nu\alpha\beta}\varepsilon_{\mu'\nu'\alpha'\beta'}\varepsilon_{\bar \mu\bar \nu\bar\alpha\bar\beta}\epsilon^{\nu}(p)
%S_{\boldsymbol{\rho}}^{\nu'\beta}(q_2)S_{\boldsymbol{K^*}}^{\alpha\bar\beta}(q_1){S_{\boldsymbol{\rho^*}}}^{\beta'\bar \nu}(q),\nonumber\\
%%()
&&\mathcal{M}_b=\int \frac{d^4q}{(2\pi)^4} g_{{\scriptscriptstyle K^{\!*}\! K^{\!*}\!\pi}}g_{{\scriptscriptstyle K^{\!*}\! K^{\!*}\!\rho}} g_{{\scriptscriptstyle TVV}} ({p_2}+q)_{\mu}{q_1}_{\mu'}q_{m}p^{\alpha}\varepsilon_{\alpha\beta\gamma\delta}\varepsilon^{\mu'm\rho\sigma}\epsilon^{*}_{\nu}(p_2)\epsilon^{\beta}(p)
S_{\boldsymbol{\rho}}^{\mu\delta}(q_2){S_{\boldsymbol{K^*}}}^{\nu}_{\sigma}(q){S_{\boldsymbol{K^*}}}^{\gamma}_{\rho}(q_1),\nonumber\\
&&\mathcal{M}_c=\int \frac{d^4q}{(2\pi)^4} g_{{\scriptscriptstyle K^{\!*}\! K^{\!*}\!\rho}}g_{{\scriptscriptstyle K^{\!*}\! K\!\rho}} g_{{\scriptscriptstyle TVV}} \varepsilon_{\alpha\beta\gamma\delta}\varepsilon^{\mu'm\rho\sigma}q_{\rho}p^{\alpha}_{2}q_{2\mu'}(q-p_{1})_{\mu}\epsilon^{*}_{\nu}(p_1)\epsilon^{\beta}(p)
{S_{\boldsymbol{\rho}}}^{\gamma}_{m}(q_2){S_{\boldsymbol{K^*}}}_{\sigma}^{\nu}(q){S_{\boldsymbol{K^*}}}_{\mu \delta}(q_1),\nonumber\\
%&&M_c=\int \frac{d^4q}{i(2\pi)^4} g_{\pi K K^*}g_{\rho K^* K} g_{TVV}(q_{\mu}- p_{1\mu})p^{\alpha}p_{2\mu'}q_{2\rho}\varepsilon_{\alpha\beta\gamma\delta}\varepsilon^{\mu'\nu\rho\sigma}\epsilon^{\beta}(p)\epsilon^{*}_{\sigma}(p_2)
%\frac{(-g^{\mu\delta}+\frac{q_{1}^{\mu}q_{1}^{\delta}}{m_{K^*}^2})(-g^{\gamma}_{\nu}+\frac{{q_2}^{\gamma}q_{2\nu}}{m_{\rho}^2})}{(q_2^2-m_{\rho}^2)(q^2-m_q^2)(q_1^2-m_{K^*}^2)},\\
%&&M_d=\int \frac{d^4q}{i(2\pi)^4} g_{K^* K^*\rho}g_{\rho K^* K^*} g_{TVV} (q_{\mu}g_{\nu}^{\nu'}-q_{\nu}g_{\mu}^{\nu'})\epsilon^{*\nu}(p_2)\epsilon^{\beta}(p)q_{\mu'}q_{\rho}p^{\alpha}\varepsilon_{\alpha\beta\gamma\delta}\varepsilon^{\mu'm\rho\sigma}
%\frac{(-g_{\nu'\sigma}+\frac{q_{\nu'}q_{\sigma}}{m_q^2})(-g^{\mu\gamma}+\frac{q_{2}^{\mu}q_{2}^{\gamma}}{m_{K^*}^2})(-g^{\delta}_{m}+\frac{q_{1m}q_{1}^{\delta}}{m_{\rho}^2})}{(q_2^2-m_{\rho}^2)(q^2-m_q^2)(q_1^2-m_K^2)}.\\
&&\mathcal{M}_d=\int \frac{id^4q}{(2\pi)^4} g_{{\scriptscriptstyle \pi \!K\! K^{\!*}}}g_{{\scriptscriptstyle \rho \rho \pi}} g_{{\scriptscriptstyle TVV}}({p_1}-q)_{\mu}p^{\alpha}p_{2\nu}q_{2\mu'}\varepsilon_{\alpha\beta\gamma\delta}\varepsilon^{\mu'\nu\rho\sigma}\epsilon^{\beta}(p)\epsilon^{*}_{\sigma}(p_2)
{S_{\boldsymbol{\rho}}}^{\delta}_{\rho}(q_2)S_{\boldsymbol{\pi}}(q)S_{\boldsymbol{K^*}}^{\mu\gamma}(q_1).
%,\nonumber\\
%%&&M_d=g_{\pi \pi\rho}g_{TPV}(p_2^{\mu}+p_1^{\mu})\epsilon^{\alpha}(p)\frac{i(-g_{\mu\alpha}+\frac{q_{\mu}q_{\alpha}}{q^2})}{(q^2-m_q^2)} .
\end{eqnarray}
To account for the off-shell effects of the exchanged particles, we include an additional off-shell form factor
\begin{eqnarray}
\mathcal{F}=(\frac{\Lambda^2-m_E^2}{\Lambda^2-q^2})^2,
\end{eqnarray}
where $m_E$ being the exchange particle mass and $\Lambda$ being the cutoff parameter for the exchange momentum. Here we choose $\Lambda=m_E+\alpha$ with $100$ MeV$\leq\alpha\leq300$ MeV. The total decay widths of $T_{\rho K}$ and  $S_{\rho K^*}$  can reach several MeV, and will increase with the cutoff parameter $\alpha$.
\begin{eqnarray}
&\left\{\begin{array}{l} \Gamma(T_{\rho K}^{--})=2.89\ \text{MeV}, \\
 \Gamma(T_{\rho K}^{+})=2.67\ \text{MeV},
\end{array}\right.
&\left\{\begin{array}{l}\Gamma(S_{\rho K^*}^{--})=9.69\ \text{MeV},\\ \Gamma(S_{\rho K^*}^{-})=9.6\ \text{MeV}.
\end{array}\right.
\end{eqnarray}
Here, we take $\alpha=200$ MeV. The antiparticles $T_{\rho K}^{++}$($S_{\rho K}^{++}$) and $T_{\rho K}^{-}$($S_{\rho K}^{+}$) are similar and therefore not shown. For completeness, the possible two-body and three-body decay widths are given in Tab.\ref{tab:numb}. Some discussions deserve attention.
\begin{itemize}
  \item The four-quark 27 molecular state $S_{\rho K^*}$ can decay into $\pi K^*$ and $K \rho$, then we define a ratio $R_1$ as,
\begin{eqnarray}
R_{1}=\frac{\Gamma(S_{\rho K^*}\to \pi K^*)}{\Gamma(S_{\rho K^*}\to K \rho)}=1.40,
\end{eqnarray}
it increases with cutoff parameter. When $\alpha=200$ MeV, the ratio $R_1$ is $1.40$.
  \item As three-body decay, we also define a similar ratio $R_2$ and $R_2'$ for the two molecular states $T_{\rho K}$ and $S_{\rho K^*}$,
\begin{eqnarray}
&&R_{2}=\frac{\Gamma(T_{\rho K}^{--}\to K^0 \pi^-\pi^-)}{\Gamma(T_{\rho K}^{--}\to K^- \pi^0 \pi^-)}=\frac{\Gamma(T_{\rho K}^{+}\to K^- \pi^+\pi^+)}{\Gamma(T_{\rho K}^{+}\to K^0 \pi^0 \pi^+)}=2\times 10^{-4},\\
&&R_{2}'=\frac{\Gamma(S_{\rho K^*}^{--}\to \overline K^0 \pi^-\pi^-)}{\Gamma(S_{\rho K^*}^{--}\to K^- \pi^0 \pi^-)}=\frac{\Gamma(S_{\rho K^*}^{-}\to \overline K^+ \pi^-\pi^-)}{\Gamma(S_{\rho K^*}^{-}\to K^0 \pi^0 \pi^-)}=0.57.
\end{eqnarray}
The ratios $R_2$ and $R_2'$ calculated in effective Lagrangian approach are smaller than that from SU(3) flavor phenomenological analysis, which means significant SU(3) flavor symmetry breaking. Particularly the ratio $R_2$ of $T_{\rho K}$ molecular state has greater deviation,  it is likely that there is contribution from three-body tree diagram in $T_{\rho K}^{--}\to K^- \pi^0 \pi^-$ and $T_{\rho K}^{+}\to K^0 \pi^0 \pi^+$.
  \item Generally, the contribution of two-body decay is larger than that of three-body process. Nevertheless, the three-body decays $T_{\rho K}^{--}\to K^-\pi^0 \pi^-$ and $T_{\rho K}^+\to K^0\pi^0 \pi^+$ are dominant for molecular state $T_{\rho K}$.
\end{itemize}

%%%%%%%%%%%%%%%%%%%%%%%%%%%%
\begin{figure}
\includegraphics[width=0.95\columnwidth]{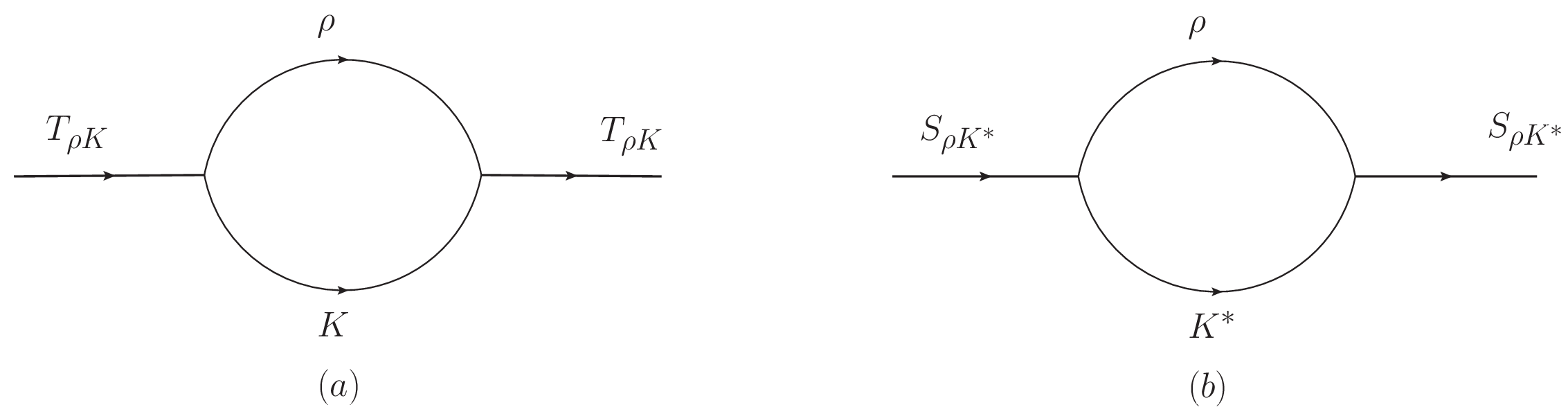}
\caption{The self-energy diagrams for four-quark 10 state($T_{\rho K}$) and 27 state($S_{\rho K^*}$).}
\label{fig:production}
\end{figure}
%%%%%%%%%%%%%%%%%%%%%%%%%%%%
%%%%%%%%%%%%%%%%%%%%%%%%%%%%
\begin{figure}
\includegraphics[width=0.95\columnwidth]{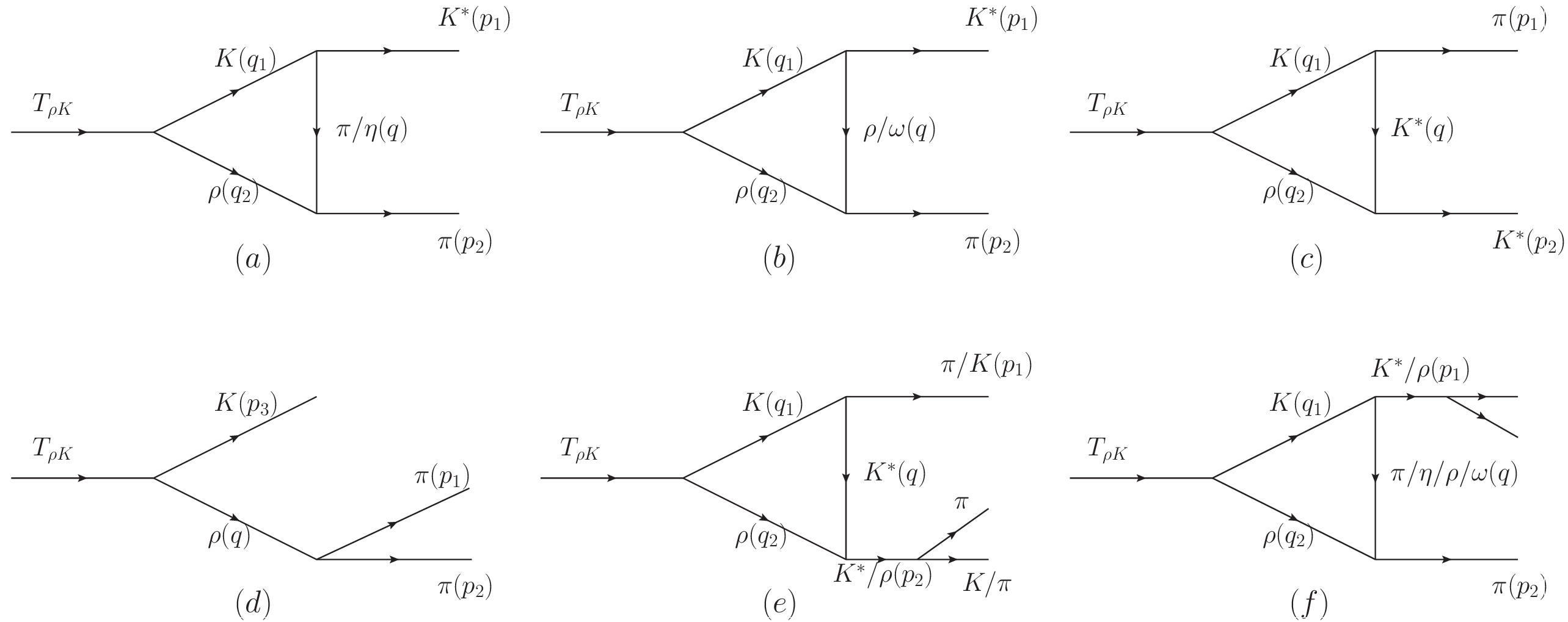}
\caption{The triangle diagrams of two body decays of four-quark decuplet in (a-c), and the diagrams of three body decays of strange four-quark 10 states in (d-f).}
\label{fig:decay}
\end{figure}
%%%%%%%%%%%%%%%%%%%%%%%%%%%%
%%%%%%%%%%%%%%%%%%%%%%%%%%%%
\begin{figure}
\includegraphics[width=0.95\columnwidth]{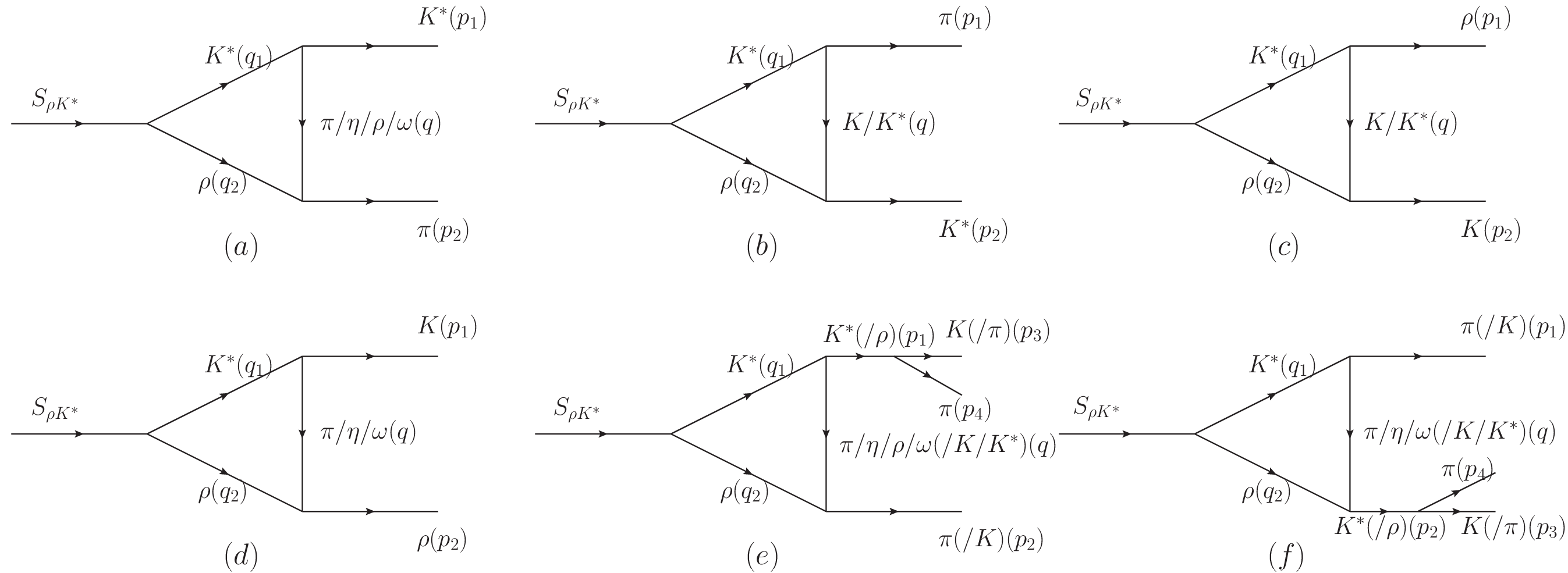}
\caption{The triangle diagrams of two body decays of four-quark 27 state in (a-d), and the diagrams of three body decays of strange four-quark 27 state in (e-f).}
\label{fig:decay2}
\end{figure}
%%%%%%%%%%%%%%%%%%%%%%%%%%%%
\begin{table}
  \centering  \caption{The two-body and three-body decay widths of strange four-quark 10 and 27 state. We adopt cutoff parameter $\alpha=100, 150, 200$ MeV respectively. }\label{tab:numb}
  \begin{tabular}{r l c c c r l c c c c c c c}
     \hline
    \multicolumn{2}{c}{\multirow{2}*{Channel}} & \multicolumn{3}{c}{Width(/MeV)} & \multicolumn{2}{c}{\multirow{2}{*}{Channel}}&\multicolumn{3}{c}{Width(/MeV)} \\ \cline{3-5} \cline{8-10}
    & & $\alpha= 100$ & $\alpha= 150$ & $\alpha= 200$ &  & & $\alpha= 100$ & $\alpha= 150$ & $\alpha= 200$ \\\hline
%    $T_{8q}\to \rho \pi$ &133 MeV &133 MeV& 133 MeV& 4.75 \\
    $T_{\rho K}^{--}$ &$\to \pi^{-} K^{*-}$ &0.14 &0.31 & 0.58 & $T_{\rho K}^{+}$ &$\to \pi^{+} K^{*0}$ &0.15 &0.32 & 0.60  \\
    &$\to K^0 \pi^-\pi^-$ &$1\times 10^{-4}$ &$3\times 10^{-4}$ & $4\times 10^{-4}$&  &$\to K^- \pi^+\pi^+$ &$2\times 10^{-4}$ &$3\times 10^{-4}$ & $5\times 10^{-4}$& \\
    %&$\to K^- K^-\overline K^0$ &5.1 &5.1 & 5.2&  &$\to K^0 K^0 K^+$ &5.1 &5.1 & 5.2& \\
    &$\to K^- \pi^0 \pi^-$&2.23 &2.27&2.31 &  &$\to K^0 \pi^0 \pi^+$ &2.01 &2.04 & 2.07& \\\hline

%    $T_{8ss}\to K^* K$ &133 MeV &133 & 133& 4.79 \\
%    $T_{10q}\to \rho \pi$& 150 MeV&133 & 133& 0.26 \\
    %$S_{\rho K^*}^{--}$&$\to \pi^- K^- $& 1.3  &5.6  & 8.7 &$S_{\rho K^*}^{-}$&$\to \pi^- K^0 $& 1.3  &5.6  & 8.7  \\
    $S_{\rho K^*}^{--}$&$\to \pi^- K^{*-}$& 1.35  &3.11 & 5.58 & $S_{\rho K^*}^{-}$&$\to \pi^- \overline K^{*0}$& 1.34  &3.10 & 5.57 & \\
    &$\to  K^- \rho^-$& 1.42  &2.53 & 4.00 & &$\to K^0 \rho^{-}$& 1.41  &2.51 & 3.98 & \\
    &$\to  \pi^-\pi^- \overline K^0$& 0.01 &0.02 & 0.04& &$\to  \pi^-\pi^- K^+$& 0.01 &0.03 & 0.07& \\
%    &$\to  K^- K^- K^0$& 0.4 &0.5 & 0.5& &$\to  K^0 K^0 K^-$& 0.4 &0.5 & 0.5& \\
    &$\to  \pi^0 \pi^- K^-$& 0.01 &0.03 &0.07& &$\to  K^0 \pi^0\pi^- $& 0.01 &0.04 & 0.12& \\
%    $T_{10ss}\to K^* \bar K$& 150 MeV &133 & 133& 4.41  \\
%    $T_{10ss}\to \rho \eta$& 150 MeV &133 & 133& 4.41  \\
%    $T_{102s}\to K^* K$& 150 MeV&133 & 133 & 20.97 \\
     \hline
   \end{tabular}
\end{table}

\section{Conclusion}
We discuss the decay modes of light strange four-quark states, based on the SU(3) flavor phenomenological analysis and effective Lagrangian approach. Some valuable decay channels and decay widths are drawn from our discussion. We found that the two kinds molecular states both have a considerable decay width. Furthermore, we conduct the analysis of two-body and three-body decay channels, offering valuable insights that could enhance theoretical understanding of dynamics and guide future experimental investigations.

  \end{document}